\documentclass[final,5p,times,twocolumn,authoryear]{elsarticle}

\usepackage{amssymb}
\usepackage{lipsum}
\usepackage{array}
\usepackage{booktabs}

\usepackage[detect-all,separate-uncertainty=true]{siunitx}[=v2]
\DeclareSIUnit\litre{L}

\usepackage[version=4]{mhchem}

\usepackage{pdfpages}

\usepackage{float}
\usepackage{newfloat}
\DeclareFloatingEnvironment[fileext=los,name=Schema,placement=htbp]{schema}
\usepackage{dblfloatfix}

\usepackage[hidelinks]{hyperref}

\begin{document}

\onecolumn

\thispagestyle{empty}

\clearpage

\setcounter{page}{1}

\begin{frontmatter}



\title{Synthesis, Characterization, and Application of Silica Aerogel Thin Films Prepared via Surface Derivatization: A Comprehensive Study on Micro-Silica Nanofilms for Enhanced Photoelectric Conversion Efficiency in Photovoltaic Solar Cells}


\author[first]{Yan Che}
\affiliation[first]{organization={Loomis Chaffee School},%Department and Organization
            addressline={4 Batchelder Rd},
            city={Windsor},
            postcode={06095},
            state={CT},
            country={USA}}

\begin{abstract}
Silica aerogels are nanoporous materials with exceptional optical and physical properties, making them promising candidates to enhance solar cell efficiency as antireflective coatings. This study synthesized hydrophobic silica aerogel thin films under ambient conditions and characterized their porous structure, surface morphology, and optical performance. The films were deposited on monocrystalline silicon solar cells to assess their impact on photovoltaic properties. A two-step acid/base catalyzed sol-gel process was utilized, followed by solvent exchange and surface modification with trimethylchlorosilane. Structural analysis via SEM revealed successful deposition of crack-free films when aging occurred in an ethanol environment. The aerogel displayed considerable specific surface area (115 m${}^{2}$/g), porosity (77.92\%) and surface roughness (55-78\%) along with a low refractive index (1.05), benefiting light harvesting. Preliminary solar testing showed increased output voltage with a 0.2 mm aerogel coating versus a bare cell. Further IV measurements demonstrated enhanced charge transport and conversion efficiency for the treated cell. The antireflective and light-trapping effects of aerogel appear to improve photon absorption. This initial research validates the potential of ambient pressure-synthesized hydrophobic silica aerogels to increase the performance of silicon photovoltaics cost-effectively. Further optimization of film thickness and morphology could realize higher efficiency gains.
\end{abstract}



\begin{keyword}
%% keywords here, in the form: keyword \sep keyword, up to a maximum of 6 keywords
silica aerogel thin film \sep monocrystalline silicon solar cell \sep ambient pressure \sep surface derivatization



\end{keyword}

\end{frontmatter}

%\tableofcontents



\section{Introduction}

Since the discovery of opalescent silica aerogels, researchers have extensively explored these highly porous materials with ultra-low density \citep{Kistler1931}. Aerogels are renowned as exceptional thermal insulators, with applications ranging from Olympic stadiums to solar receivers. For example, aerogel insulation was utilized in roofing, walls, and ice rinks during the 2022 Winter Olympics to maintain optimal indoor temperatures. Various groups have leveraged aerogel's transparency and insulation to improve solar absorption in a receiver, harnessing heat beyond \SI{265}{\degreeCelsius} for cost-effective industrial and domestic heating \citep{Yu2023,Zhao2019}. Despite aerogel's demonstrated solar applications, minimal research has explored its potential to enhance photoelectric conversion in solar cells. Like solar thermal receivers, solar cells absorb sunlight to generate energy. An aerogel coating could potentially improve light harvesting and conversion efficiency as an effective anti-reflective layer. Further research is warranted to translate aerogel's insulating advantages into gains for photovoltaic (PV) solar energy.

In many regions, PV technology has emerged as an economical source of electricity. The core objective of ongoing research in this field is to diminish the cost-to-efficiency ratio of solar cells. Despite the dominance of silicon (\ce{Si}) in the PV market, it is not considered the most suitable material for absorbing light. Therefore, a noteworthy approach to reducing the cost per watt involves optimizing the utilization of light that is not absorbed in its initial passage through the device. Various architectural concepts have been advanced to meet this objective. In the context of monocrystalline Si cells, one design employing random-pyramid surface textures stands out for its path-length improvement, while also being relatively straightforward to implement, such as with alkaline solutions \citep{Kocak2023}. It is essential that these designs intended to increase path length be harmonized with a high degree of internal reflectance on the absorber surfaces to curtail the loss of energy at each interaction with the light.

In this research, I aim to improve solar cell efficiency by replacing the conventional antireflective front layer with a silica aerogel thin film. Aerogels are highly desirable for this application due to their high porosity, opalescence, and air-filled structure. Silica aerogels are characteristically known to exhibit a low refractive index, approximately 1.00, coupled with high porosity, around 99\%. Furthermore, they are often observed to provide high visible light transmission, nearing 99\%, and a substantial specific surface area, typically on the order of 1000 square meters per gram \citep{Mekonnen2021}. These properties allow more light to penetrate into the aerogel while minimizing reflection at the interface. The nanoporous structure acts as an effective antireflective coating, reducing the number of photons reflected from the solar cell surface. Utilizing a silica aerogel thin film could potentially enhance solar cell efficiency without requiring modifications to the semiconductor itself.

Silica aerogels have conventionally been produced through the application of supercritical drying techniques \citep{Gurav2010}. This standard approach brings with it various challenges related to economic feasibility, consistent process flow, and flammability of solvents, primarily due to the requirements for elevated temperature and pressure to reach the crucial phase transition point \citep{Dorcheh2008,Goryunova2023}. Employing liquid carbon dioxide in a chilled supercritical drying procedure could incrementally reduce the aerogels' atmospheric chemical stability, given the intrinsic hydrophilic nature of aerogel particles.

To overcome existing challenges, Brinker et al. formulated a method for fabricating silica aerogel at ambient pressure. As shown in Schema~\ref{sch:1}, this novel methodology chemically alters the wet gel's surface by replacing hydrophobic functional groupings for hydrogen in hydroxyl groups, followed by drying at room pressure. The process triggers the adjacent silica clusters' surface silanol groups (Si--OH) to engage in condensation reactions, resulting in a permanent contraction of the gel's structure during the drying phase \citep{Yang1997}. Through the formation of extremely low-energy surfaces, this technique substantially minimizes surface tension. As a result, it becomes vital to tailor alcogel surfaces using appropriate agents, thus making the aerogel's surface water-repellent and resolving the previously noted issues \citep{Henning1981}.

The key purpose of preparing aerogel-like \ce{SiO2} thin films under ambient pressure in this work is to primarily investigate their effects on solar cell efficiency, contributing to the broader goal of not only enhancing the light absorption capabilities but also optimizing the photonic management of solar cells. This exploration, aligned with the global push towards energy sustainability, seeks to provide a potentially transformative solution for the economical and efficient utilization of solar energy, with particular attention to mitigating losses associated with light reflection and absorption inefficiency.

\begin{schema}[!ht]
  \centering
  \includegraphics[width=\linewidth]{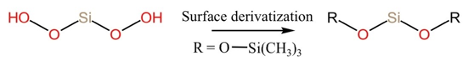}
  \caption[Surface derivatization]{Surface derivatization on replacing \ce{-OH} group with hydrophobic \ce{-O-Si(CH3)3} group. Trimethylchlorosilane (TMCS) is known to interact with both water and the hydroxyl groups found on the exterior and interior surfaces of silica gel \citep{Zhang2001}. These reactions can be depicted through the subsequent chemical expressions: \\
  \centerline{\ce{2(CH3)3SiCl + H_2O <--> (CH3)3Si-O-Si(CH3)3 + 2HCl};} \\
  \centerline{\ce{(CH3)3SiCl + HO-Si# -> (CH3)3Si-O-Si + HCl}.}}
  \label{sch:1}
  \vspace*{-2ex}
\end{schema}

\section{Method}

\subsection{Thin film preparation}

Silica aerogel samples were synthesized utilizing two analogous recipes, distinguished primarily by the drying processes employed. The first method omits a drying process in an ethanol environment, while the second incorporates this step and demonstrates increased time efficiency. In the latter method, the film is deposited onto the surface prior to drying. The former technique represents the patented ambient pressure method as proposed by Brinker et al., while the latter has been refined by subsequent researchers. Within the time constraint of this study, only the scanning electron microscopy (SEM) results of the aerogels prepared using the modified method are presented, with the underlying rationales elucidated in subsequent discussions:

\begin{enumerate}
\item  The preparation of silicate sols involved the dissolution of tetraethoxysilane (TEOS) in ethanol (EtOH) through a two-step acidic-basic catalyzed procedure (termed B2 solution) \citep{Brinker1999}. Initially, TEOS, EtOH, \ce{H2O}, and \ce{HCl} were combined in the molar ratios of 1.0:3.8:1.1:\num{7.0e-4}, and this mixture was refluxed at \SI{60}{\degreeCelsius} for a duration of 90 minutes, leading to the formation of a stock solution. Subsequently, a mixture of 0.05M \ce{NH4OH} stock solution, EtOH, and \ce{H2O}, in volume ratios of 1:10:44, was prepared, resulting in \SI{55}{\milli\litre} of B2 solution. This solution was then allowed to gel and age at \SI{60}{\degreeCelsius}. A typical \SI{2}{\milli\metre} thin film gel necessitated approximately 46 hours for gelation and a minimum of 92 hours for drying at \SI{60}{\degreeCelsius}. For surface derivatization, washing steps were performed at \SI{50}{\degreeCelsius} and included three rinses in EtOH, two in n-hexane, followed by a 20-hour soak in a silylating solution of 5 vol\% TMCS and 95 vol\% n-hexane, and an additional rinse in n-hexane. Ultimately, the reliquification of the gel was executed by placing the coated monocrystalline silicon panel in an ultrasound machine (\SI{20}{\kilo\hertz} frequency, \SI{475}{\watt} maximum power), complemented by additional n-hexane and 20 minutes of sonication. The gels were subsequently applied to a mono-crystalline silicon panel utilizing a microscopic slide.
\item  A two-step acid/base catalyzed method was used to prepare the \ce{SiO2} sol \citep{Yang1999}. First, tetraethoxysilane (TEOS, 99.5\%, Shanghai Macklin Biochemical Technology Co., Ltd.), ethanol (EtOH, 99.7\%, Macklin), water, and hydrochloric acid (\ce{HCl}, 37\% by mass, Macklin) were combined in a 1:3:1:\num{1.4e-3} molar ratio and mixed at room temperature for 90 minutes. Next, a mixture of \SI{14}{\milli\litre} ethanol and \SI{2}{\milli\litre} 0.05 M ammonium hydroxide (\ce{NH4OH}, 25\% by mass, Shanghai Adamas Reagent Co., Ltd.) was added to \SI{10}{\milli\litre} of the stock solution and mixed for 30 minutes \citep{Prakash1995a}. The resulting wet gel was brushed onto a \SI{125x125}{\milli\metre} p-doped monocrystalline silicon solar panel (MSL Solar Energy Co. Ltd.) using a microscopic slide. The coated substrate was then placed in an ethanol vapor-saturated container to strengthen the wet gel film network structure through an aging process \citep{Yang1999}. To facilitate subsequent surface derivatization, the pore fluids in the wet gel film were exchanged with ethanol and n-hexane ($>99.0\%$, Macklin) sequentially (TMCS ($>98.0\%$, Macklin) does not react with n-hexane, in which it has good miscibility). These solvent exchanges were performed by immersing the coated substrate in the respective solvent at \SI{60}{\degreeCelsius} for 12 hours. The substrate was subsequently immersed in a 6 vol\% TMCS/n-hexane solution at \SI{60}{\degreeCelsius} for 12 hours to proceed with the derivatization process. Following this step, the modified wet gel film was rinsed in n-hexane for a period of 12 hours, then dried at \SI{60}{\degreeCelsius}. The subsequent heating process was carried out at \SI{300}{\degreeCelsius}, with a heating rate of \SI{1}{\degreeCelsius} per minute and a soaking time of 2 hours, followed by a cooling rate of \SI{2}{\degreeCelsius} per minute.
\end{enumerate}

\subsection{Characterization}

Fourier-transform infrared spectroscopy (FTIR, ThermoNicolet 6700, wave-number range \SIrange{400}{4000}{\per\centi\metre}, resolution ratio \SI{48}{\per\centi\metre}, scan number: 32) was carried out to distinguish the properties of the silica films. Scanning electron microscopy (SEM, FEI Quanta 650FEG) was employed to study the pore size and morphology of the silica films. Samples were affixed to a conductive adhesive surface to avoid contamination, and then sputter-coated with gold, a procedure necessitated by the high dielectric constant of silica aerogel, which does not image well without an electric conducting coating under SEM \cite{Hotaling1993,Hrubesh1994,Lin2023}. Observations were conducted in high vacuum mode, using an accelerating voltage of 10 kV and magnifications of \SIrange{1}{2}{\micro\metre}. Atomic force microscopy (AFM) was employed to assess the surface roughness of the thin films, allowing precise measurements at the nanoscale level. Mercury intrusion porosimetry (MIP, MicroActive AutoPore V 9600) was conducted to measure the porosity of the aerogel, providing essential information regarding the internal structure and pore volume. Brunauer -- Emmett -- Teller (BET, Micromeritics ASAP 2460) surface area analysis was conducted to supplement the porosity analysis. This approach facilitated a more comprehensive understanding of the surface and porous characteristics of the aerogel, elucidating both the microstructure and the specific surface area that are essential to the material's performance and properties.

\subsection{Solar cell setup}

In the evaluation of solar cell efficiency, a simulated terrestrial environment was established, conforming to AM 1.5 standards and maintained at a temperature of \SI{25}{\degreeCelsius}. A lamp (Philips 66027, 27W) was strategically positioned 20 cm above the \SI{125x125}{\milli\metre} solar panel to ensure an orthogonal incidence of light upon the surface. Preliminary measurements were undertaken with a multimeter (UNI-T UT33A+), connected in parallel with the solar cell, to ascertain the output voltage of both untreated and treated solar cells as shown in Schema~\ref{sch:2}. To safeguard the electrical apparatus within the circuit, a variable resistor (\SIrange{0}{5}{\ohm}) was incorporated in series with the solar cell.

\begin{schema}[!ht]
  \centering
  \includegraphics[scale=0.67]{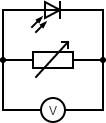}
  \caption{Circuit diagram for preliminary measurement of solar cell output voltage.}
  \label{sch:2}
\end{schema}

After these preliminary measurements, a \SI{3x5}{\milli\metre} fragment was excised from the treated and untreated solar panel using a single-point diamond dresser. An IV curve was generated with a Source Measure Unit (Keithley's Standard Series 2400 SMU), accompanied by a Nanovoltmeter (Keithley Model 2182A) as shown in Schema~\ref{sch:3}. The preliminary data guided the determination of the maximal voltage applicable to the solar cell, minimizing the risk of electrical breakdown in the P-N junction.

\begin{schema}[!ht]
  \centering
  \includegraphics[scale=0.67]{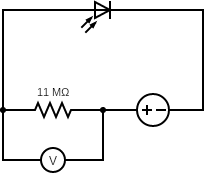}
  \caption{Circuit diagram for IV curve measurement of solar cell: DC supply as Keithley 2400, and the voltmeter as Keithley 2182A.}
  \label{sch:3}
\end{schema}

Upon initial measurement of the internal resistance of the sample using a multimeter (which was found to be around \SI{7}{\mega\ohm}), a resistor of similar resistance was selected. However, it was rapidly observed during the measurement that the voltage drawn by the multimeter when assessing the sample's resistance exceeded the maximum operational voltage of the sample piece, triggering electrical breakdown. Subsequent investigations into this phenomenon, including a reassessment of resistance (where the solar cell swiftly transitioned into a good conductor with little resistance about \SI{16}{\ohm} in a dark environment, forfeiting its semiconducting property), revealed the occurrence of Avalanche breakdown. This resulted in a state where the junction could not revert to its original position, culminating in the complete destruction of the diode.

\section{Results and Discussion}

\subsection{Properties of silica aerogel thin films}

\begin{figure*}[!ht]
  \centering
  \includegraphics[width=0.8\linewidth]{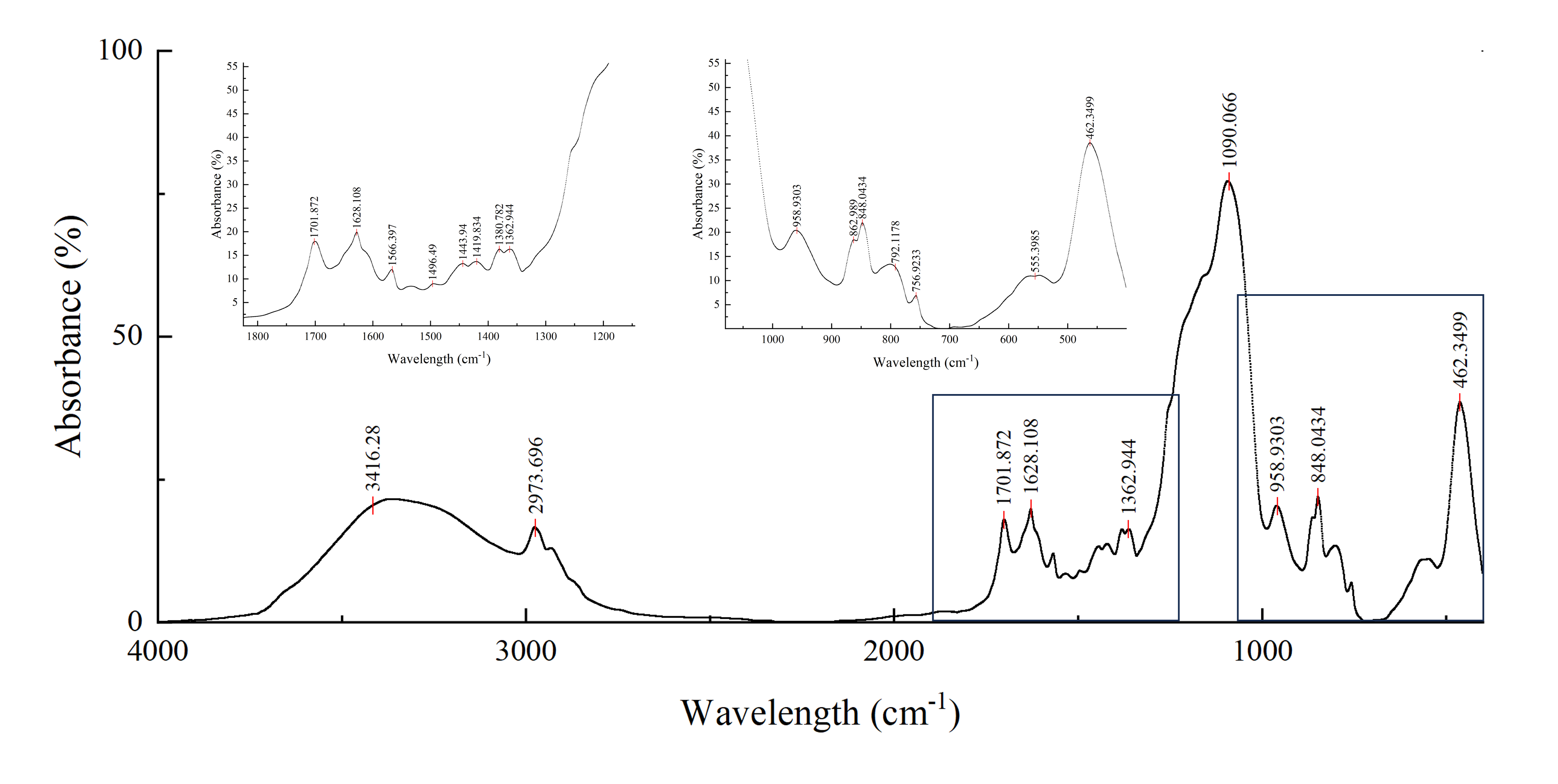}
  \caption{FTIR spectrum of silica thin film.}
  \label{fig:1}
\end{figure*}

FTIR spectrum of aerogel thin film is shown in Fig.~\ref{fig:1}, revealing several distinctive peaks, which reflects the complex structural characteristics of the material. The broadest peak at \SI{3416}{\per\centi\metre} is indicative of O-H stretching vibrations, likely associated with residual silanol (\ce{Si-OH}) groups within the structure. This indicates the moisture absorption in the film, which may be attributed to the incomplete water removal in the hydrolysis and polycondensation of the sol-gel reaction \citep{Kawakami2000}. These may represent unreacted hydroxyl functionalities that have not been fully converted through derivatization. Peaks at \SI{2973}{\per\centi\metre} are consistent with \ce{C-H} stretches of methyl groups, potentially signifying the presence of \ce{-O-Si(CH3)3} groups formed through the reaction with TMCS \citep{Acquaroli2016,Zhang2001}. The prominent peak at \SI{1090}{\per\centi\metre} corresponds to Si-O-Si stretches, highlighting the inherent siloxane backbone of the silica network \citep{Budunoglu2011,Yang1997,Zhang2001}. Additional peaks at \SI{1701}{\per\centi\metre}, \SI{1628}{\per\centi\metre}, and \SI{1240}{\per\centi\metre} could be attributed to various functional groups introduced or modified during the derivatization process, possibly including carbonyl or ether functionalities \citep{Nandiyanto2019}. The lower wavenumber peaks, such as those at \SI{555}{\per\centi\metre} and \SI{462}{\per\centi\metre} (the second-highest peak), might reflect specific bending or rocking motions within the silica structure.

\begin{figure}[!ht]
  \centering
  \includegraphics[width=\linewidth]{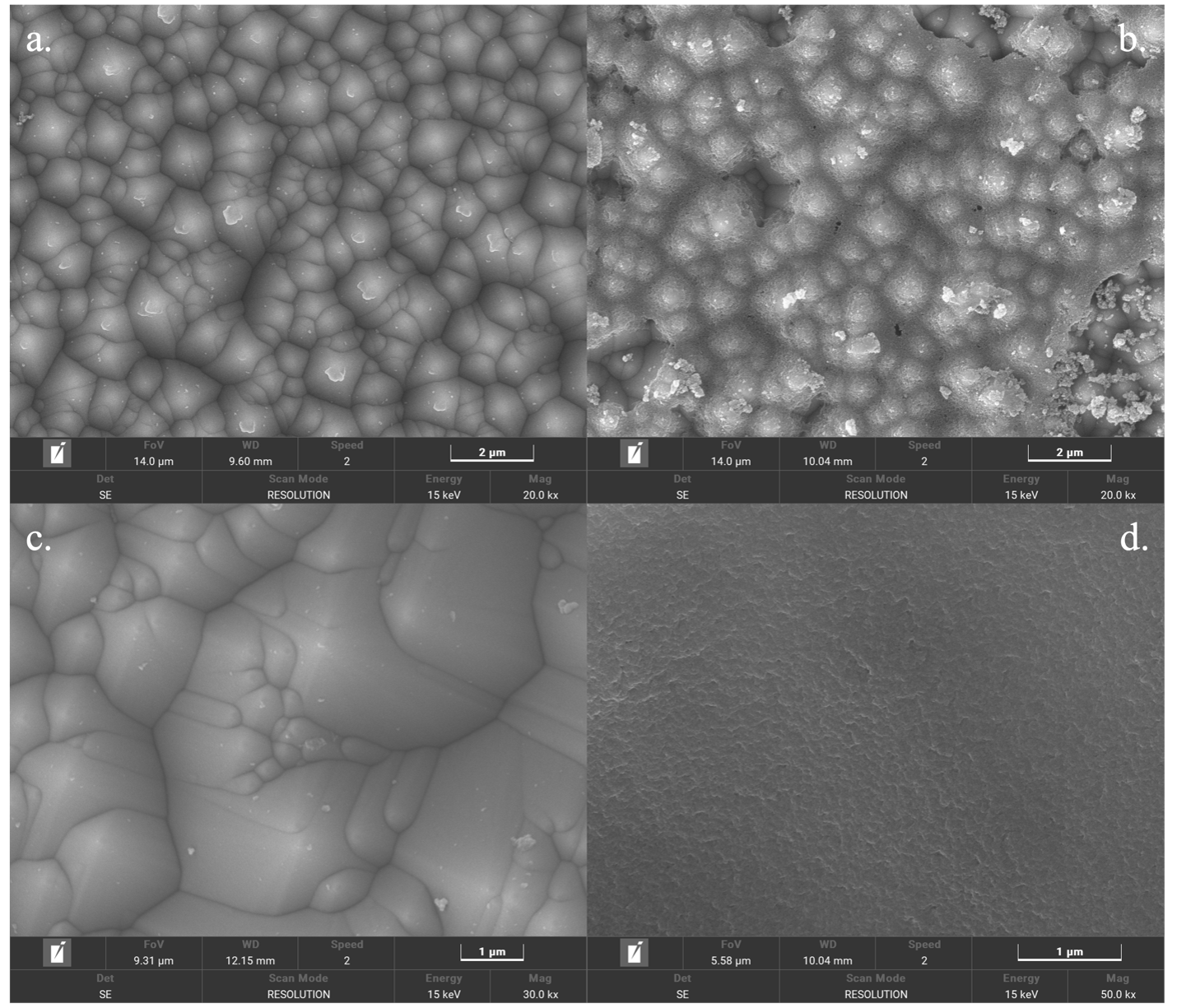}
  \caption{SEM images of solar cells: (a), (c) as-prepared solar cell without any film deposition; (b) solar cell treated with surface derivatization steps, excluding an ethanol environment; (d) solar cell treated with surface derivatization steps, conducted specifically in an ethanol environment.}
  \label{fig:2}
\end{figure}

Fig.~\ref{fig:2} presents SEM scans of the solar cell, illustrating significant details related to the structure and deposition process. As observed in sections (a) and (c), the unit cell of monocrystalline silicon appears in good condition. In contrast, section (b) reveals an apparent thin film, detectable through slight image blurring. Interestingly, the film does not completely obscure the underlying solar panel, allowing for visualization of the conducting phase beneath. This observation suggests that the film has not been uniformly deposited onto the solar cell. Such transparency within SEM imaging is inconsistent with the expected properties of a high-dielectric material, indicating a misalignment in the deposition process. Further examination of (b) shows a poorly structured film with numerous holes and fractures. This points to a phenomenon of volume shrinkage due to capillary pressure ($P_{c}$) during ambient drying, likely attributable to the lack of an ethanol environment during its aging \citep{Kim2004}. The absence of ethanol during aging may lead to the silicon network structure collapsing when the alcoholic substances are evaporated, resulting in structural defects. This shrinkage, observable during the drying process, is predominantly determined by two main factors: the capillary pressure within the pore fluid and the rigidity of the inorganic gel structure \citep{Prakash1995b}. Capillary pressure is a function of several factors, including the liquid/vapor surface tension ($\sigma_{LV}$), the contact angle ($\theta$) of the liquid on the pore wall, and the radius of the pore ($r_{p}$). Assuming cylindrical pores, the capillary pressure can be represented by a specific equation, commonly expressed using the Young-Laplace equation:

\[P_c=\frac{2\sigma_{LV}}{r_m}=\frac{2{\sigma }_{LV}{\mathrm{cos} \theta \ }}{r_p}\]

A noteworthy observation pertains to the surface tensions of the drying solvents, specifically n-hexane, which exhibits a $\sigma_{LV}$ of \SI{18.2}{\milli\newton\per\metre} at \SI{20}{\degreeCelsius} (Klein et al., 2019). When compared to the work conducted by Kim et al.'s group, who successfully prepared silica aerogel with high porosity and high surface area morphology utilizing isopropanol (\SI{21.7}{\milli\newton\per\metre} at \SI{20}{\degreeCelsius}) and n-heptane (\SI{20.14}{\milli\newton\per\metre} at \SI{20}{\degreeCelsius}) as drying solvents, the $\sigma_{LV}$ of n-hexane is evidently lower under identical conditions for preparing wet gel films \citep{Kim2004}. Nevertheless, wet gel films dried from n-hexane may experience an increase in capillary pressure relative to those dried from isopropanol and n-heptane. This phenomenon can be attributed to variations in the contact angles of the drying solvents on the silica network, impacting the overall drying process and the resultant aerogel properties.

Conversely, the aerogel prepared in an ethanol environment, as seen in (d), exhibits a uniformly distributed film across the surface, with no detectable fragments or holes. The concealed solar panel and wrinkle-like features in the image align with the intended film structure, reflecting the presence of a silicon network skeleton. The stark contrast between (c) and (d) underscores the pivotal role of ethanol in the aerogel's aging process, highlighting its importance in maintaining structural integrity and preventing volume shrinkage.

\begin{figure}[!ht]
  \centering
  \includegraphics[width=\linewidth]{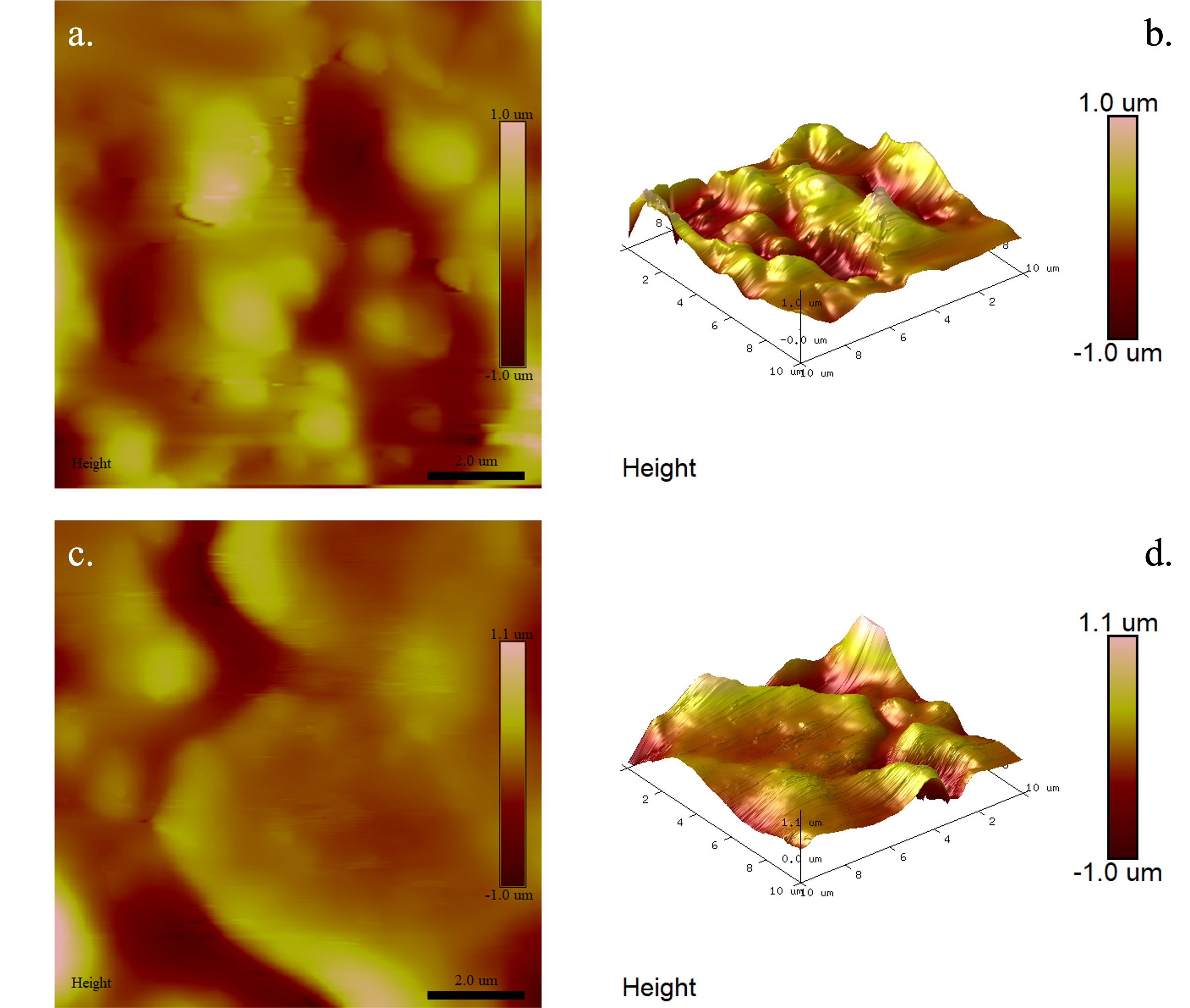}
  \caption{AFM height scan of aerogel thin film. (a) and (b) depict the 2D and 3D scans, respectively, over a specific \SI{10x10}{\micro\metre} region; (c) and (d) represent corresponding 2D and 3D scans of another distinct region.}
  \label{fig:3}
\end{figure}

AFM was employed to investigate the surface morphology of the aerogel thin film, as illustrated in Fig.~\ref{fig:3}. Through the utilization of NanoScope Analysis software, the analysis of regions (a) and (b) revealed a median surface roughness of 55.83\% and an average roughness of 62.89\%. Conversely, regions (c) and (d) exhibited a median roughness of 62.32\% and an average roughness of 78.83\%. These quantitative measurements were further accentuated by the recorded surface depth (the distance from the highest to lowest peak) of \SI{1.935}{\micro\metre} in regions (a) and (b), and \SI{1.407}{\micro\metre} in regions (c) and (d).

The observed roughness data signifies a highly textured surface, suggestive of an ideal candidate for replacing conventional random-pyramid surfaces. Such a rough surface can contribute to an increased optical path length, thus enhancing the material's light-harvesting capability. Furthermore, the substantial surface depth indicates the potential of the material to trap incident light that initially reflects off the surface. This characteristic, coupled with the complex surface morphology, can serve to minimize surface reflection, thereby optimizing light utilization.

The variance in roughness and depth measurements between the two regions indicates the existence of heterogeneous characteristics across the thin film. Such heterogeneity could be indicative of localized variations in the deposition or synthesis process, contributing to differing morphological features. This spatial variation in surface properties could further influence the optical and mechanical performance of the aerogel, affecting parameters such as reflection, absorption, and mechanical resilience. Consequently, future work may involve the development of more versatile synthesis and deposition procedures to understand and control these surface features. This could lead to tailored optimization of the aerogel thin film for specific photovoltaic applications, maximizing efficiency and functionality within solar cell systems.

\begin{figure*}[!ht]
  \centering
  \includegraphics[width=0.75\linewidth]{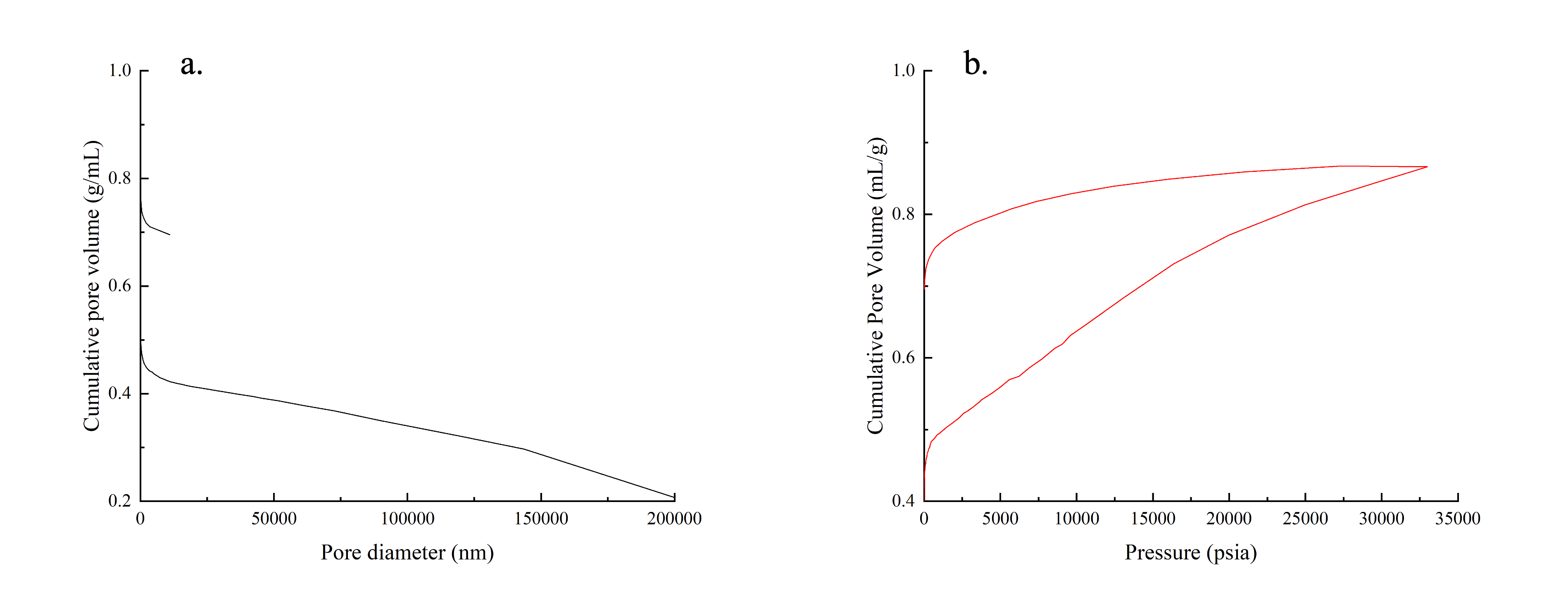}
  \caption{Mercury intrusion curves for silica aerogel thin film.}
  \label{fig:4}
\end{figure*}

MIP results is shown in Table.1 and Fig.~\ref{fig:4}. A high total intrusion volume of \SI{0.8661}{\milli\litre\per\gram} and pore area of \SI{115.001}{\square\metre\per\gram} underscores the extensive internal surface area. The disparity between the median pore diameter (volume) at \SI{6305.61}{\nano\metre} and the median pore diameter (area) at \SI{9.39}{\nano\metre} points to a complex distribution of pore sizes, indicative of the hierarchical porous nature typical to aerogels. An average pore diameter of \SI{30.13}{\nano\metre} complements these findings. The bulk density of \SI{0.2752}{\gram\per\milli\litre}, coupled with an apparent density of \SI{1.25}{\gram\per\milli\litre}, reflects the film's lightweight characteristics. The porosity could thus be calculated as:
\[\Pi = \frac{\rho_a-\rho_b}{\rho_a} \times 100\]
Where $\rho_{a}$ is the apparent density and $\rho_{b}$ is the bulk density.

The porosity is notably high at 77.92\%, with an interstitial porosity of 47.63\%, highlighting the film's open structure, which would be too porous for reflective optics \citep{Hotaling1993}. The permeability of 484.85 mdarcy indicates a well-connected pore structure that would allow for the efficient passage of fluids, potentially enhancing diffusion or conduction of photons within the aerogel. The tortuosity factor of 1.643 in the silica aerogel thin film indicates a complex pore structure, leading to an increased scattering of light within the material. This enhanced scattering effect can extend the path of light, thereby contributing to greater light absorbance, a feature that may be beneficial in applications such as photovoltaic solar cells. These features collectively suggest a material with significant potential for enhancing light absorption through increased path length and reduced reflection, a key consideration in improving photovoltaic solar cell efficiency.

\begin{table*}[!ht]
  \centering
  \caption{Mercury intrusion data for silica aerogel thin film.}
  \label{tab:1}
  \small
  \begin{tabular}{>{\bfseries}cc}
    \toprule
    Total intrusion volume at 32,994.66 psia:                                              & \SI{0.8661}{\milli\litre\per\gram}   \\
    Total pore area at 32,994.66 psia:                                                     & \SI{115.001}{\square\metre\per\gram} \\
    Median pore diameter (volume) at 28.68 psia and \SI{0.433}{\milli\litre\per\gram}:     & \SI{6305.61}{\nano\metre}            \\
    Median pore diameter (area) at 19,261.35 psia and \SI{57.500}{\square\metre\per\gram}: & \SI{9.39}{\nano\metre}               \\
    Average pore diameter (4V/A):                                                          & \SI{30.13}{\nano\metre}              \\
    Bulk density at 0.50 psia:                                                             & \SI{0.2752}{\gram\per\milli\litre}   \\
    Apparent (skeletal) density at 32,994.66 psia:                                         & \SI{1.2466}{\gram\per\milli\litre}   \\
    Porosity:                                                                              & 77.92\%                              \\
    \midrule
    Permeability:                                                                          & 484.8547 mdarcy                      \\
    Threshold pressure:                                                                    & 9.30 psia (Calculated)               \\
    Characteristic length:                                                                 & \SI{19446.40}{\nano\metre}           \\
    Conductivity formation factor:                                                         & 0.290 (Calculated)                   \\
    Tortuosity factor:                                                                     & 1.643                                \\
    Tortuosity:                                                                            & 29.0387                              \\
    Percolation fractal dimension:                                                         & 2.858                                \\
    Backbone fractal dimension:                                                            & 2.962                                \\
    \midrule
    Interstitial porosity:                                                                 & 47.63\%                              \\
    Breakthrough pressure ratio:                                                           & 3.3512                               \\
    Linear coefficient:                                                                    & -3.0670E-05 1/psia                   \\
    Quadratic coefficient:                                                                 & 5.3925E-10 1/psia\textsuperscript{2} \\
    Calculated porosity:                                                                   & 77.92\%                              \\
    \bottomrule
  \end{tabular}
\end{table*}

\begin{figure*}[!ht]
  \centering
  \includegraphics[width=0.8\linewidth]{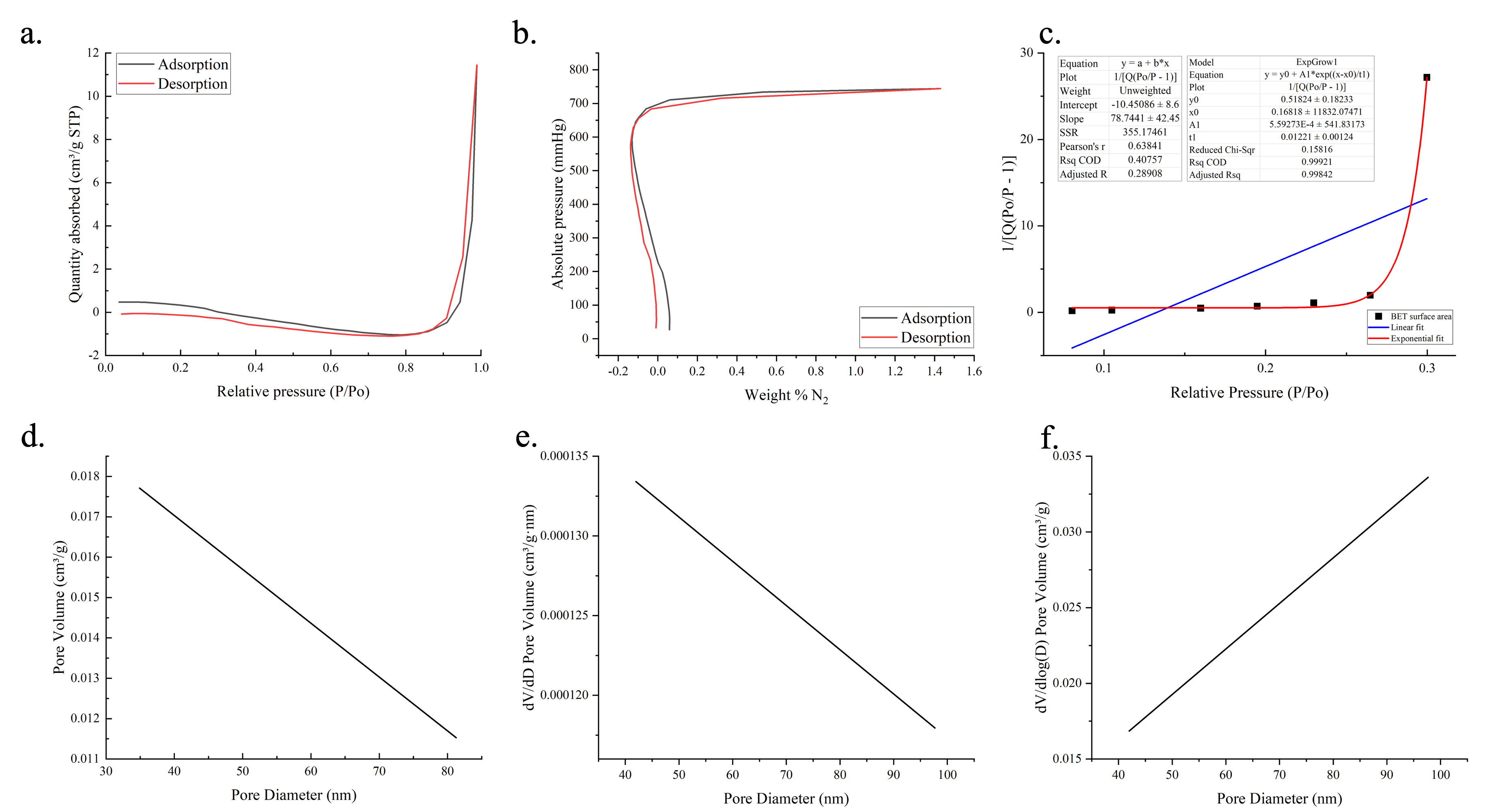}
  \caption{BET data of silica aerogel thin film..}
  \label{fig:5}
\end{figure*}

BET data is displayed in Fig.~\ref{fig:5}. In (a), from an initial increase in gas quantity adsorbed as the pressure increases (from $\sim$0.035 to $\sim$0.265 relative pressure), a consistent decrease in the quantity adsorbed is observed until a relative pressure of approximately 0.945, at which point, the trend reverses, and the quantity adsorbed dramatically increases. The increase in gas adsorption at lower pressures indicates the availability of surface sites for adsorption and might be indicative of the mesoporous structure of the aerogel. According to the classification by the International Union of Pure and Applied Chemistry (IUPAC), silica aerogels exhibited type IV isotherms and H3 hysteresis loops within the relative pressure range of $P/P_{0}=0.8-1.1$, indicative of the presence of numerous mesopores (Xie et al., 2022). However, the sudden change in trend with an extreme increase in the quantity adsorbed at higher relative pressures represents an unusual behavior, possibly indicative of experimental errors or unique structural characteristics of the aerogel. Such inconsistencies could also point to a multi-layer adsorption phenomenon or specific interactions between the gas molecules and the aerogel surface \citep{Goryunova2023}. The tortuosity factor previously discussed is also likely to play a role here, as the complex pore structure can significantly affect the path and interaction of gas molecules with the surface. The detailed morphology, pore distribution, and surface characteristics of the silica aerogel would require further examination and experimental validation to derive a more comprehensive understanding of this trend. As observed in (b), the absence of a simple mirror pattern between the adsorption and desorption curves indeed indicates that the process is not purely reversible. This could be attributed to kinetic constraints and the presence of metastable states in the system. In the context of the silica aerogel thin film, the tortuous pore structure may lead to varying diffusion rates of the adsorbate into and out of the pores, causing a deviation from simple reversibility. Additionally, the specific interactions between the adsorbate and the pore walls, along with the potential formation of metastable states during adsorption or desorption, might contribute to this complex pattern. Such features emphasize the uniqueness of the adsorption-desorption process in silica aerogels, reflecting a dynamic interplay between kinetic factors, pore geometry, and molecular interactions. Consequently, in (c), reveals a surface area of \SI{0.0637 \pm 0.0404}{\square\metre\per\gram} and a monolayer capacity Qm of \SI{0.0146}{\cubic\centi\metre}/STP, indicating a limited and potentially irregular surface area. The slope and Y-intercept values, along with a negative C constant, may suggest non-ideal adsorption behavior or a lack of monolayer formation. The correlation coefficient of 0.638 further points to a moderate fit to the BET model, which, combined with the sharp increase in $1/[Q(P_{0}/P-1)]$ at higher relative pressures, raises questions about the specific nature of adsorption in this material. Given these inconsistencies and the meso- to mega-sized pore structure of the silica aerogel, reliance on mercury intrusion data may be more appropriate for characterization. BET measurements might not be sensitive enough for this specific pore size range, rendering mercury intrusion a more trustworthy method for assessing the material's porous characteristics \cite{Fricke1992}.

Silica aerogels frequently result in opalescent materials in many production procedures, a trait that is atypical for porous substances. This uncommon amalgamation of characteristics is derived from the aerogel's microstructure, possessing a scale smaller than the wavelength of visible light. The phenomenon of opalescence is engendered by minimal scattering within the visible spectrum, wherein the scattered light sustains an isotropic angular distribution and exhibits restricted multiple scattering \citep{Beck1989}. Such a phenomenon aligns with the principles of Rayleigh scattering theory, characterized by isotropic scattering of vertically polarized incident light. The intensity of the scattering manifests as $\cos^{2}\theta$ for horizontally polarized incident light, with a scattering that is wavelength-dependent and inversely proportional to the fourth power of the wavelength \citep{Beck1989,Piegari1985}. Despite the alignment with Rayleigh scattering, aerogels differ from perfect Rayleigh scatterers; they may also manifest a component of scattering that is wavelength-independent and may not be isotropic, with some samples potentially deviating substantially from the typical Rayleigh angular distribution \citep{Beck1989}.

The refractive index (n) of the aerogels was determined using Clausius--Mossotti relation and Lorentz--Lorenz equation \citep{Budunoglu2011}:

\[n=\frac{3}{2}\frac{{n_s}^2-1}{{n_s}^2+2}\frac{\rho }{{\rho }_s}=0.19\rho +1=1.05\]

In the given equation, n and $n_{s}$ represent the refractive indices of the aerogel and silica, respectively, while the corresponding densities for these materials are denoted by $\rho$ and $\rho_{s}$. By utilizing values of $n_{s}=1.46$ for the refractive index of silica and a density of $\rho_{s} = \SI{2.2}{\gram\per\cubic\centi\metre}$, the refractive index of the aerogel can be calculated as $n=1+0.19\rho$.

\begin{table*}[!ht]
  \centering
  \caption{Preliminary measurements of solar cell output voltage. $V_{oc}$ represents the open-circuit voltage, and latter columns states the output power of solar cell connected with difference resistance. Vertical axis denotes the various thicknesses of the thin film. The thickness of each film is calculated by dividing the volume of the bulk aerogel by the area of the silicon panel.}
  \label{tab:2}
  \small
  \begin{tabular}{ccccccc}
    \toprule
                           & $V_{oc}$ (mV) & $P_{5\Omega }$ (mW) & $P_{4\Omega}$ (mW) & $P_{3\Omega}$ (mW) & $P_{2\Omega}$ (mW) & $P_{1\Omega}$ (mW) \\
    \midrule
                           & 457           & 16.70              & 15.63              & 14.01              & 11.25              & 7.92               \\
    w/o film               & 450           & 16.36              & 16.13              & 15.12              & 12.01              & 8.10               \\
                           & 452           & 15.68              & 14.16              & 13.74              & 11.70              & 9.22               \\
    \midrule
                           & 468           & 19.47              & 19.60              & 15.99              & 11.70              & 9.03               \\
    \SI{0.2}{\milli\metre} & 463           & 19.47              & 19.32              & 15.55              & 11.10              & 9.41               \\
                           & 466           & 19.22              & 20.16              & 16.13              & 12.48              & 9.03               \\
    \midrule
                           & 450           & 17.17              & 17.56              & 14.70              & 10.80              & 8.46               \\
    \SI{0.4}{\milli\metre} & 460           & 17.29              & 17.42              & 15.41              & 11.10              & 9.22               \\
                           & 458           & 16.82              & 17.29              & 15.41              & 9.66               & 7.06               \\
    \midrule
                           & 428           & 14.05              & 14.76              & 12.55              & 5.62               & 5.78               \\
    \SI{0.8}{\milli\metre} & 436           & 14.47              & 14.52              & 11.04              & 6.27               & 4.90               \\
                           & 430           & 13.83              & 13.81              & 11.53              & 6.61               & 4.62               \\
    \bottomrule
  \end{tabular}
\end{table*}

The silica aerogel thin film's refractive index of 1.05, which is remarkably close to that of air, contributes to its excellent optical properties. This low refractive index indicates minimal refraction and dispersion of light as it passes through the aerogel, allowing a high transmission of light and giving rise to the material's characteristic transparency. The unique porous microstructure of the aerogel results in a weighted average of the refractive indices of the silica and air, leading to this low value. Consequently, the low refractive index permits light to penetrate the material with little reflection or scattering, a quality that makes the aerogel suitable for various optical applications, alluding to an ideal candidate for antireflective coatings in photovoltaic solar cells. Nevertheless, the utilization of a base-catalyzed tetramethyl orthosilicate (TMOS) formula has been documented to exhibit superior optical characteristics, exhibiting ninefold reduced scattering, and concomitant minimal contraction within the gel architecture, which suggests a refractive index more akin to that of air \citep{Russo1986}. The disparity in transparency is not solely attributable to variations in particle or pore size, as evidenced by the BET measurements, which indicate the presence of larger particles within TMOS gels \citep{Pisal2016}. Future investigations must be undertaken to identify the most efficacious sol recipe for the synthesis of aerogels with elevated transparency.

\subsection{Solar cell efficiency testing}

The preliminary testing data of the solar cell, as shown in Table~\ref{tab:2}, reveals significant insights into the impact of silica aerogel thin film thickness on output power. As the thickness of the thin film increases from \SIrange{0.0}{0.8}{\milli\metre}, a generally decreasing trend in the output power is observed across all resistance values. This trend could suggest a relationship between the thin film's thickness and the solar cell's efficiency, with thicker films possibly impeding photon flow. However, a noteworthy observation is the comparison between no film and the \SI{0.2}{\milli\metre} film. The data shows a significant increment in the output voltage across all resistances for the \SI{0.2}{\milli\metre} film, indicating that the silica aerogel thin film indeed influences the solar cell in a manner that increases its output power. This observation highlights the potential benefits of silica aerogel thin films in enhancing photovoltaic performance, matching with Hrubesh et al.'s discussion regarding commercial use of thick aerogel films on the cover slip of solar cell, though the optimal thickness requires further investigation \citep{Hrubesh1995}. Additionally, the peak output power was observed when the solar cell was connected to a \SI{4}{\ohm} resistor, indicating that the internal resistance of the solar cell approximated \SI{4}{\ohm}.

According to the preliminary data, the maximum output voltage was found to be less than 500 mV. Consequently, during the IV measurement, the applied voltage should be substantially lower than this maximum output voltage. This consideration is necessary as the testing sample represents a significantly smaller portion of the entire solar panel, resulting in the samples exhibiting greater resistance and a reduced maximum working voltage.

\begin{figure*}[!ht]
  \centering
  \includegraphics[width=0.8\linewidth]{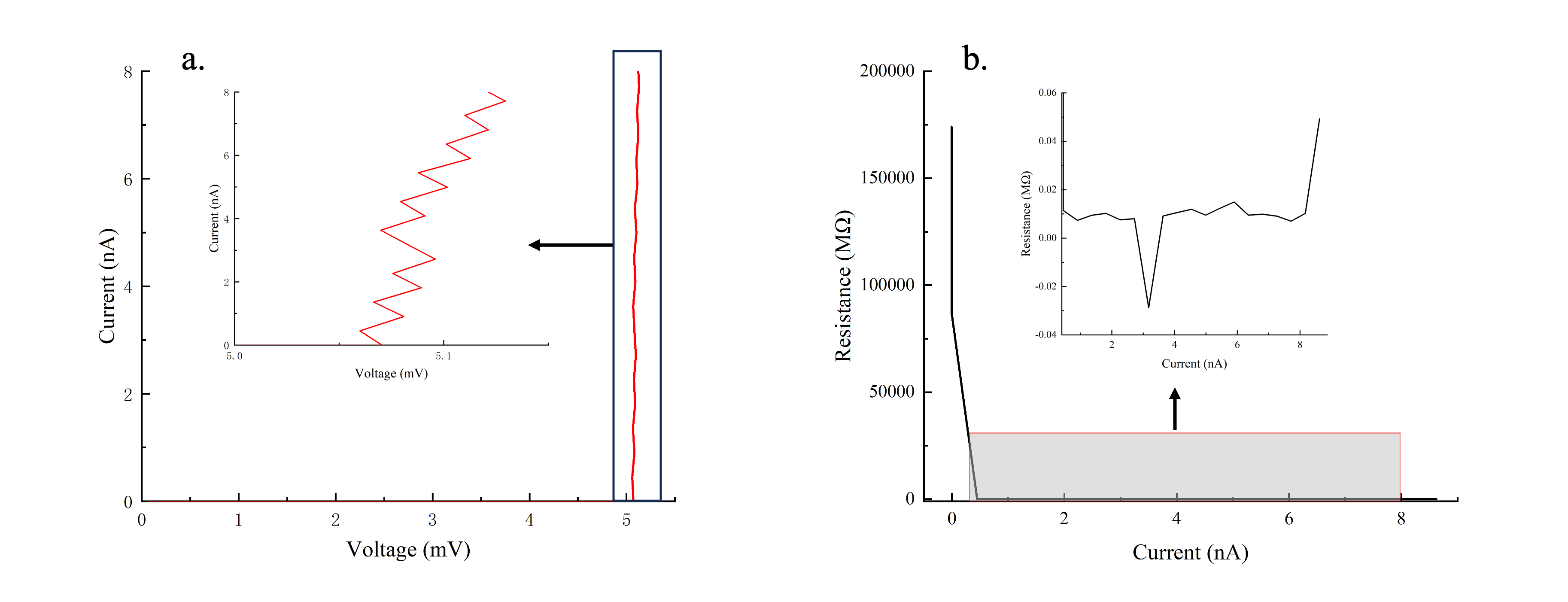}
  \caption{Electrical transportation measurement of solar cell in dark environment: (a) IV curve with voltage running from 0-100mV; (b) IR curve of the same set up by taking the derivative of the IV curve.}
  \label{fig:6}
\end{figure*}

The results of the primary electrical transport measurements of the solar cell are depicted in Fig.~\ref{fig:6}. In part (a), the current primarily remains at zero as the voltage increases, indicative of an initial insulating property. Upon reaching its maximum threshold voltage, electrical breakdown occurs within the diode, and the solar cell begins to exhibit conductive properties, a characteristic behavior of an ideal semiconductor. The corresponding IR curve, shown in part (b), further elucidates this phenomenon. Initially, the resistance is substantial, consistent with insulating behavior. Subsequently, there is a significant and abrupt decrease in resistance, which then approaches a plateau, suggesting a transition from insulator to conductor.

\begin{figure*}[!ht]
  \centering
  \includegraphics[width=\linewidth]{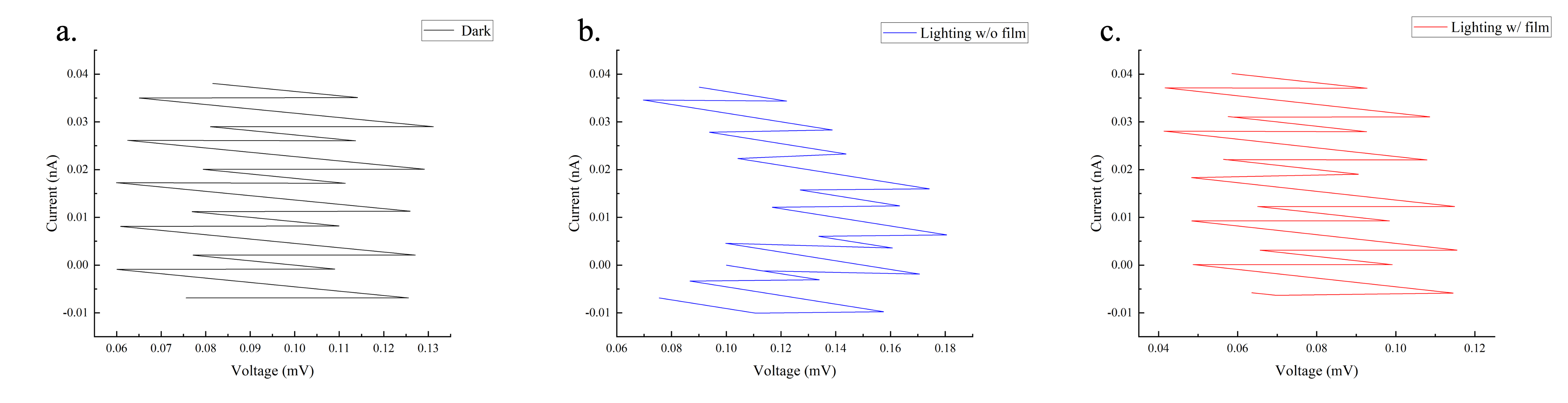}
  \caption{IV curves of solar cell: (a) As-prepared solar cell tested in a dark environment; (b) As-prepared solar cell tested in a lighting environment; (c) Treated solar cell tested in a lighting environment.}
  \label{fig:7}
\end{figure*}

The IV curves of the as prepared and treated solar cells are depicted in Fig.~\ref{fig:7}, providing insight into their distinct electrical behaviors. Upon close examination and comparison of curves (a) and (c), a noticeable leftward horizontal deviation in the curve is discernible for the treated solar cell. As the current increases, there is a gradual decrement in the current reading within the solar cell coated with aerogel thin film. This phenomenon suggests that the treated solar cell has generated an intrinsic voltage that counteracts the applied voltage. In contrast, curve (b) exhibits a more erratic behavior, swaying left and right without a discernible pattern. This random fluctuation implies that the as-prepared solar cell, when tested under a lighting condition, produces an unstable output voltage. The lack of stability in the output voltage might be indicative of certain inconsistencies in the material or fabrication process, which warrants further investigation.

The comparison between the as-prepared and treated solar cells highlights the impact of the aerogel thin film coating. The treated solar cell's more controlled response to varying current levels suggests an enhancement in the photovoltaic properties, leading to higher photoelectric conversion efficiency.  The leftward sway observed in the treated solar cell curve could be indicative of a better-controlled charge transport mechanism, potentially reducing recombination losses and enhancing the extraction of photogenerated carriers. Moreover, the aerogel's unique nanostructured matrix may play a vital role in optimizing the light-trapping properties, thereby contributing to the improved conversion efficiency.

\section{Conclusion}

In conclusion, this study presents compelling evidence that a silica aerogel thin film can significantly enhance the photovoltaic efficiency of a monocrystalline silicon solar cell. The silica aerogel was successfully synthesized via a two-step acid/base catalyzed sol-gel process, followed by critical surface modification with TMCS under ambient conditions. Extensive characterization revealed a highly porous, lightweight material with low refractive index, high visible light transmission, and substantial surface roughness. These unique optical and morphological properties make the aerogel ideal as an antireflective coating to improve light harvesting.

Preliminary solar cell testing showed an increase in output voltage with a \SI{0.2}{\milli\metre} aerogel film compared to an untreated cell. Further IV curve analysis demonstrated enhanced photoelectric conversion for the treated cell, with controlled charge transport behavior. The aerogel likely minimizes surface reflections while increasing optical path length within the cell via its textured interface. Overall, the nanoscale structural complexity provides optimal light trapping to boost photon absorption.

While thicker aerogel films were found to reduce output voltage, this initial study clearly validates the potential of silica aerogels to improve silicon-based photovoltaics. With further optimization of thickness and deposition, as well as surface functionalization, even higher efficiency gains could be attained. Additionally, the scalable synthetic procedure developed here enables large-scale, low-cost fabrication, overcoming prior economic barriers.

In the broader context, this research makes a compelling case for pursuing and integrating silica aerogels within next-generation solar technologies. The approach leverages an ingenious material to mitigate fundamental limitations and deficiencies of existing photovoltaics. As ongoing work enhances fabrication methods and expands the aerogel's photoactive properties, exceptionally high-performance yet inexpensive solar cells are envisioned. The insights from this study thus constitute a significant advance toward innovative solar designs for meeting global energy demands in a sustainable and renewable manner.

\section*{Acknowledgement}
\addcontentsline{toc}{section}{Acknowledgement}

I wish to express my sincere gratitude to Mr. Osei-Mensah for his invaluable support, without which this work would not have been possible. I would also like to thank Dr. Sarah Maurer and Dr. Stephanie Bertenshaw from Central Connecticut State University for graciously allowing me access to their laboratory over break to fabricate samples. Furthermore, I am deeply appreciative of Yuqin Wu from Brown University for her guidance throughout this research endeavor. Finally, I would like to extend my thanks to Qiang Zhao and Wenlong Yang from Beijing Normal University, who provided me with excellent characterization resources that greatly facilitated this work.

\clearpage

\bibliographystyle{elsarticle-harv}
\bibliography{refs}

\begin{thebibliography}{29}
\expandafter\ifx\csname natexlab\endcsname\relax\def\natexlab#1{#1}\fi
\providecommand{\url}[1]{\texttt{#1}}
\providecommand{\href}[2]{#2}
\providecommand{\path}[1]{#1}
\providecommand{\DOIprefix}{doi:}
\providecommand{\ArXivprefix}{arXiv:}
\providecommand{\URLprefix}{URL: }
\providecommand{\Pubmedprefix}{pmid:}
\providecommand{\doi}[1]{\href{http://dx.doi.org/#1}{\path{#1}}}
\providecommand{\Pubmed}[1]{\href{pmid:#1}{\path{#1}}}
\providecommand{\bibinfo}[2]{#2}
\ifx\xfnm\relax \def\xfnm[#1]{\unskip,\space#1}\fi
%Type = Article
\bibitem[{Acquaroli et~al.(2016)Acquaroli, Newby, Santato and
  Peter}]{Acquaroli2016}
\bibinfo{author}{Acquaroli, L.N.}, \bibinfo{author}{Newby, P.},
  \bibinfo{author}{Santato, C.}, \bibinfo{author}{Peter, Y.A.},
  \bibinfo{year}{2016}.
\newblock \bibinfo{title}{Thermal properties of methyltrimethoxysilane aerogel
  thin films}.
\newblock \bibinfo{journal}{{AIP} Advances} \bibinfo{volume}{6}.
\newblock \DOIprefix\doi{10.1063/1.4965921}.
%Type = Article
\bibitem[{Beck et~al.(1989)Beck, Caps and Fricke}]{Beck1989}
\bibinfo{author}{Beck, A.}, \bibinfo{author}{Caps, R.},
  \bibinfo{author}{Fricke, J.}, \bibinfo{year}{1989}.
\newblock \bibinfo{title}{Scattering of visible light from silica aerogels}.
\newblock \bibinfo{journal}{Journal of Physics D: Applied Physics}
  \bibinfo{volume}{22}, \bibinfo{pages}{730--734}.
\newblock \DOIprefix\doi{10.1088/0022-3727/22/6/002}.
%Type = Misc
\bibitem[{Brinker and Prakash(1999)}]{Brinker1999}
\bibinfo{author}{Brinker, C.J.}, \bibinfo{author}{Prakash, S.S.},
  \bibinfo{year}{1999}.
\newblock \bibinfo{title}{Ambient pressure process for preparing aerogel thin
  films reliquified sols useful in preparing aerogel thin films}.
\newblock \bibinfo{note}{US Patent 5,948,482}.
%Type = Article
\bibitem[{Budunoglu et~al.(2011)Budunoglu, Yildirim, Guler and
  Bayindir}]{Budunoglu2011}
\bibinfo{author}{Budunoglu, H.}, \bibinfo{author}{Yildirim, A.},
  \bibinfo{author}{Guler, M.O.}, \bibinfo{author}{Bayindir, M.},
  \bibinfo{year}{2011}.
\newblock \bibinfo{title}{Highly transparent, flexible, and thermally stable
  superhydrophobic {ORMOSIL} aerogel thin films}.
\newblock \bibinfo{journal}{{ACS} Applied Materials {\&} Interfaces}
  \bibinfo{volume}{3}, \bibinfo{pages}{539--545}.
\newblock \DOIprefix\doi{10.1021/am101116b}.
%Type = Article
\bibitem[{Dorcheh and Abbasi(2008)}]{Dorcheh2008}
\bibinfo{author}{Dorcheh, A.S.}, \bibinfo{author}{Abbasi, M.H.},
  \bibinfo{year}{2008}.
\newblock \bibinfo{title}{Silica aerogel; synthesis, properties and
  characterization}.
\newblock \bibinfo{journal}{Journal of Materials Processing Technology}
  \bibinfo{volume}{199}, \bibinfo{pages}{10--26}.
\newblock \DOIprefix\doi{10.1016/j.jmatprotec.2007.10.060}.
%Type = Incollection
\bibitem[{Fricke and Emmerling(1992)}]{Fricke1992}
\bibinfo{author}{Fricke, J.}, \bibinfo{author}{Emmerling, A.},
  \bibinfo{year}{1992}.
\newblock \bibinfo{title}{Aerogels{\textemdash}preparation, properties,
  applications}, in: \bibinfo{booktitle}{Chemistry, Spectroscopy and
  Applications of Sol-Gel Glasses}. \bibinfo{publisher}{Springer Berlin
  Heidelberg}, pp. \bibinfo{pages}{37--87}.
\newblock \DOIprefix\doi{10.1007/bfb0036965}.
%Type = Article
\bibitem[{Goryunova et~al.(2023)Goryunova, Gahramanli and
  Gurbanova}]{Goryunova2023}
\bibinfo{author}{Goryunova, K.}, \bibinfo{author}{Gahramanli, Y.},
  \bibinfo{author}{Gurbanova, R.}, \bibinfo{year}{2023}.
\newblock \bibinfo{title}{Adsorption properties of silica aerogel-based
  materials}.
\newblock \bibinfo{journal}{{RSC} Advances} \bibinfo{volume}{13},
  \bibinfo{pages}{18207--18216}.
\newblock \DOIprefix\doi{10.1039/d3ra02462h}.
%Type = Article
\bibitem[{Gurav et~al.(2010)Gurav, Jung, Park, Kang and Nadargi}]{Gurav2010}
\bibinfo{author}{Gurav, J.L.}, \bibinfo{author}{Jung, I.K.},
  \bibinfo{author}{Park, H.H.}, \bibinfo{author}{Kang, E.S.},
  \bibinfo{author}{Nadargi, D.Y.}, \bibinfo{year}{2010}.
\newblock \bibinfo{title}{Silica aerogel: synthesis and applications}.
\newblock \bibinfo{journal}{Journal of Nanomaterials} \bibinfo{volume}{2010},
  \bibinfo{pages}{1--11}.
\newblock \DOIprefix\doi{10.1155/2010/409310}.
%Type = Article
\bibitem[{Henning and Svensson(1981)}]{Henning1981}
\bibinfo{author}{Henning, S.}, \bibinfo{author}{Svensson, L.},
  \bibinfo{year}{1981}.
\newblock \bibinfo{title}{Production of silica aerogel}.
\newblock \bibinfo{journal}{Physica Scripta} \bibinfo{volume}{23},
  \bibinfo{pages}{697--702}.
\newblock \DOIprefix\doi{10.1088/0031-8949/23/4b/018}.
%Type = Article
\bibitem[{Hotaling(1993)}]{Hotaling1993}
\bibinfo{author}{Hotaling, S.P.}, \bibinfo{year}{1993}.
\newblock \bibinfo{title}{Ultra-low density aerogel optical applications}.
\newblock \bibinfo{journal}{Journal of Materials Research} \bibinfo{volume}{8},
  \bibinfo{pages}{352--355}.
\newblock \DOIprefix\doi{10.1557/jmr.1993.0352}.
%Type = Incollection
\bibitem[{Hrubesh and Pekala(1994)}]{Hrubesh1994}
\bibinfo{author}{Hrubesh, L.W.}, \bibinfo{author}{Pekala, R.W.},
  \bibinfo{year}{1994}.
\newblock \bibinfo{title}{Dielectric properties and electronic applications of
  aerogels}, in: \bibinfo{booktitle}{Sol-Gel Processing and Applications}.
  \bibinfo{publisher}{Springer {US}}, pp. \bibinfo{pages}{363--367}.
\newblock \DOIprefix\doi{10.1007/978-1-4615-2570-7_31}.
%Type = Article
\bibitem[{Hrubesh and Poco(1995)}]{Hrubesh1995}
\bibinfo{author}{Hrubesh, L.W.}, \bibinfo{author}{Poco, J.F.},
  \bibinfo{year}{1995}.
\newblock \bibinfo{title}{Thin aerogel films for optical, thermal, acoustic and
  electronic applications}.
\newblock \bibinfo{journal}{Journal of Non-Crystalline Solids}
  \bibinfo{volume}{188}, \bibinfo{pages}{46--53}.
\newblock \DOIprefix\doi{10.1016/0022-3093(95)00028-3}.
%Type = Article
\bibitem[{Kawakami et~al.(2000)Kawakami, Fukumoto, Kinoshita, Suzuki and
  Inoue}]{Kawakami2000}
\bibinfo{author}{Kawakami, N.}, \bibinfo{author}{Fukumoto, Y.},
  \bibinfo{author}{Kinoshita, T.}, \bibinfo{author}{Suzuki, K.},
  \bibinfo{author}{Inoue, K.i.}, \bibinfo{year}{2000}.
\newblock \bibinfo{title}{Preparation of highly porous silica aerogel thin film
  by supercritical drying}.
\newblock \bibinfo{journal}{Japanese Journal of Applied Physics}
  \bibinfo{volume}{39}, \bibinfo{pages}{L182}.
\newblock \DOIprefix\doi{10.1143/jjap.39.l182}.
%Type = Article
\bibitem[{Kim and Hyun(2004)}]{Kim2004}
\bibinfo{author}{Kim, G.S.}, \bibinfo{author}{Hyun, S.H.},
  \bibinfo{year}{2004}.
\newblock \bibinfo{title}{Synthesis and characterization of silica aerogel
  films for inter-metal dielectrics via ambient drying}.
\newblock \bibinfo{journal}{Thin Solid Films} \bibinfo{volume}{460},
  \bibinfo{pages}{190--200}.
\newblock \DOIprefix\doi{10.1016/j.tsf.2003.12.151}.
%Type = Article
\bibitem[{Kistler(1931)}]{Kistler1931}
\bibinfo{author}{Kistler, S.S.}, \bibinfo{year}{1931}.
\newblock \bibinfo{title}{Coherent expanded aerogels and jellies}.
\newblock \bibinfo{journal}{Nature} \bibinfo{volume}{127},
  \bibinfo{pages}{741--741}.
\newblock \DOIprefix\doi{10.1038/127741a0}.
%Type = Article
\bibitem[{Kocak et~al.(2023)Kocak, Tsoi, Turkay, Koc, Donercark, Budunoglu and
  Yerci}]{Kocak2023}
\bibinfo{author}{Kocak, D.}, \bibinfo{author}{Tsoi, K.},
  \bibinfo{author}{Turkay, D.}, \bibinfo{author}{Koc, M.},
  \bibinfo{author}{Donercark, E.}, \bibinfo{author}{Budunoglu, H.},
  \bibinfo{author}{Yerci, S.}, \bibinfo{year}{2023}.
\newblock \bibinfo{title}{Silica aerogel as rear reflector in silicon
  heterojunction solar cells for improved infrared response}.
\newblock \bibinfo{journal}{Solar Energy Materials and Solar Cells}
  \bibinfo{volume}{258}, \bibinfo{pages}{112430}.
\newblock \DOIprefix\doi{10.1016/j.solmat.2023.112430}.
%Type = Article
\bibitem[{Lin et~al.(2023)Lin, Mah, Randi, DeFrances, Bernot and
  Talghader}]{Lin2023}
\bibinfo{author}{Lin, P.}, \bibinfo{author}{Mah, M.}, \bibinfo{author}{Randi,
  J.}, \bibinfo{author}{DeFrances, S.}, \bibinfo{author}{Bernot, D.},
  \bibinfo{author}{Talghader, J.J.}, \bibinfo{year}{2023}.
\newblock \bibinfo{title}{High average power optical properties of silica
  aerogel thin film}.
\newblock \bibinfo{journal}{Thin Solid Films} \bibinfo{volume}{768},
  \bibinfo{pages}{139722}.
\newblock \DOIprefix\doi{10.1016/j.tsf.2023.139722}.
%Type = Article
\bibitem[{Mekonnen et~al.(2021)Mekonnen, Ding, Liu, Guo, Pang, Ding and
  Seid}]{Mekonnen2021}
\bibinfo{author}{Mekonnen, B.T.}, \bibinfo{author}{Ding, W.},
  \bibinfo{author}{Liu, H.}, \bibinfo{author}{Guo, S.}, \bibinfo{author}{Pang,
  X.}, \bibinfo{author}{Ding, Z.}, \bibinfo{author}{Seid, M.H.},
  \bibinfo{year}{2021}.
\newblock \bibinfo{title}{Preparation of aerogel and its application progress
  in coatings: a mini overview}.
\newblock \bibinfo{journal}{Journal of Leather Science and Engineering}
  \bibinfo{volume}{3}.
\newblock \DOIprefix\doi{10.1186/s42825-021-00067-y}.
%Type = Article
\bibitem[{Nandiyanto et~al.(2019)Nandiyanto, Oktiani and
  Ragadhita}]{Nandiyanto2019}
\bibinfo{author}{Nandiyanto, A.B.D.}, \bibinfo{author}{Oktiani, R.},
  \bibinfo{author}{Ragadhita, R.}, \bibinfo{year}{2019}.
\newblock \bibinfo{title}{How to read and interpret {FTIR} spectroscope of
  organic material}.
\newblock \bibinfo{journal}{Indonesian Journal of Science and Technology}
  \bibinfo{volume}{4}, \bibinfo{pages}{97}.
\newblock \DOIprefix\doi{10.17509/ijost.v4i1.15806}.
%Type = Article
\bibitem[{Piegari and Masetti(1985)}]{Piegari1985}
\bibinfo{author}{Piegari, A.}, \bibinfo{author}{Masetti, E.},
  \bibinfo{year}{1985}.
\newblock \bibinfo{title}{Thin film thickness measurement: A comparison of
  various techniques}.
\newblock \bibinfo{journal}{Thin Solid Films} \bibinfo{volume}{124},
  \bibinfo{pages}{249--257}.
\newblock \DOIprefix\doi{10.1016/0040-6090(85)90273-1}.
%Type = Article
\bibitem[{Pisal and Rao(2016)}]{Pisal2016}
\bibinfo{author}{Pisal, A.A.}, \bibinfo{author}{Rao, A.V.},
  \bibinfo{year}{2016}.
\newblock \bibinfo{title}{Comparative studies on the physical properties of
  {TEOS}, {TMOS} and {Na2SiO3} based silica aerogels by ambient pressure drying
  method}.
\newblock \bibinfo{journal}{Journal of Porous Materials} \bibinfo{volume}{23},
  \bibinfo{pages}{1547--1556}.
\newblock \DOIprefix\doi{10.1007/s10934-016-0215-y}.
%Type = Article
\bibitem[{Prakash et~al.(1995a)Prakash, Brinker and Hurd}]{Prakash1995a}
\bibinfo{author}{Prakash, S.S.}, \bibinfo{author}{Brinker, C.J.},
  \bibinfo{author}{Hurd, A.J.}, \bibinfo{year}{1995}a.
\newblock \bibinfo{title}{Silica aerogel films at ambient pressure}.
\newblock \bibinfo{journal}{Journal of Non-Crystalline Solids}
  \bibinfo{volume}{190}, \bibinfo{pages}{264--275}.
\newblock \DOIprefix\doi{10.1016/0022-3093(95)00024-0}.
%Type = Article
\bibitem[{Prakash et~al.(1995b)Prakash, Brinker, Hurd and Rao}]{Prakash1995b}
\bibinfo{author}{Prakash, S.S.}, \bibinfo{author}{Brinker, C.J.},
  \bibinfo{author}{Hurd, A.J.}, \bibinfo{author}{Rao, S.M.},
  \bibinfo{year}{1995}b.
\newblock \bibinfo{title}{Silica aerogel films prepared at ambient pressure by
  using surface derivatization to induce reversible drying shrinkage}.
\newblock \bibinfo{journal}{Nature} \bibinfo{volume}{374},
  \bibinfo{pages}{439--443}.
\newblock \DOIprefix\doi{10.1038/374439a0}.
%Type = Article
\bibitem[{Russo and Hunt(1986)}]{Russo1986}
\bibinfo{author}{Russo, R.E.}, \bibinfo{author}{Hunt, A.J.},
  \bibinfo{year}{1986}.
\newblock \bibinfo{title}{Comparison of ethyl versus methyl sol-gels for silica
  aerogels using polar nephelometry}.
\newblock \bibinfo{journal}{Journal of Non-Crystalline Solids}
  \bibinfo{volume}{86}, \bibinfo{pages}{219--230}.
\newblock \DOIprefix\doi{10.1016/0022-3093(86)90490-4}.
%Type = Article
\bibitem[{Yang et~al.(1999)Yang, Choi, Hyun and Park}]{Yang1999}
\bibinfo{author}{Yang, H.S.}, \bibinfo{author}{Choi, S.Y.},
  \bibinfo{author}{Hyun, S.H.}, \bibinfo{author}{Park, C.G.},
  \bibinfo{year}{1999}.
\newblock \bibinfo{title}{Ambient-dried {SiO} 2 aerogel thin films and their
  dielectric application}.
\newblock \bibinfo{journal}{Thin Solid Films} \bibinfo{volume}{348},
  \bibinfo{pages}{69--73}.
\newblock \DOIprefix\doi{10.1016/s0040-6090(99)00016-4}.
%Type = Article
\bibitem[{Yang et~al.(1997)Yang, Choi, Hyun, Park and Hong}]{Yang1997}
\bibinfo{author}{Yang, H.S.}, \bibinfo{author}{Choi, S.Y.},
  \bibinfo{author}{Hyun, S.H.}, \bibinfo{author}{Park, H.H.},
  \bibinfo{author}{Hong, J.K.}, \bibinfo{year}{1997}.
\newblock \bibinfo{title}{Ambient-dried low dielectric {SiO}2 aerogel thin
  film}.
\newblock \bibinfo{journal}{Journal of Non-Crystalline Solids}
  \bibinfo{volume}{221}, \bibinfo{pages}{151--156}.
\newblock \DOIprefix\doi{10.1016/s0022-3093(97)00335-9}.
%Type = Article
\bibitem[{Yu et~al.(2023)Yu, Huang, Wang, Tang and Du}]{Yu2023}
\bibinfo{author}{Yu, X.}, \bibinfo{author}{Huang, M.}, \bibinfo{author}{Wang,
  X.}, \bibinfo{author}{Tang, G.H.}, \bibinfo{author}{Du, M.},
  \bibinfo{year}{2023}.
\newblock \bibinfo{title}{Plasmon silica aerogel for improving high-temperature
  solar thermal conversion}.
\newblock \bibinfo{journal}{Applied Thermal Engineering} \bibinfo{volume}{219},
  \bibinfo{pages}{119419}.
\newblock \DOIprefix\doi{10.1016/j.applthermaleng.2022.119419}.
%Type = Article
\bibitem[{Zhang et~al.(2001)Zhang, Wang, Wu, Shen and Buddhudu}]{Zhang2001}
\bibinfo{author}{Zhang, Q.}, \bibinfo{author}{Wang, J.}, \bibinfo{author}{Wu,
  G.}, \bibinfo{author}{Shen, J.}, \bibinfo{author}{Buddhudu, S.},
  \bibinfo{year}{2001}.
\newblock \bibinfo{title}{Interference coating by hydrophobic aerogel-like
  {SiO}2 thin films}.
\newblock \bibinfo{journal}{Materials Chemistry and Physics}
  \bibinfo{volume}{72}, \bibinfo{pages}{56--59}.
\newblock \DOIprefix\doi{10.1016/s0254-0584(01)00322-4}.
%Type = Article
\bibitem[{Zhao et~al.(2019)Zhao, Bhatia, Yang, Strobach, Weinstein, Cooper,
  Chen and Wang}]{Zhao2019}
\bibinfo{author}{Zhao, L.}, \bibinfo{author}{Bhatia, B.},
  \bibinfo{author}{Yang, S.}, \bibinfo{author}{Strobach, E.},
  \bibinfo{author}{Weinstein, L.A.}, \bibinfo{author}{Cooper, T.A.},
  \bibinfo{author}{Chen, G.}, \bibinfo{author}{Wang, E.N.},
  \bibinfo{year}{2019}.
\newblock \bibinfo{title}{Harnessing heat beyond 200 {\textdegree}c from
  unconcentrated sunlight with nonevacuated transparent aerogels}.
\newblock \bibinfo{journal}{{ACS} Nano} \bibinfo{volume}{13},
  \bibinfo{pages}{7508--7516}.
\newblock \DOIprefix\doi{10.1021/acsnano.9b02976}.

\end{thebibliography}


%Type = Article

\clearpage

\appendix

\section*{Supplementary materials}
\addcontentsline{toc}{section}{Supplementary materials}

\subsection*{Preparation of \ce{SiO2} precursor with the first recipe}

A mixture of 30.5 mL tetraethyl orthosilicate (TEOS), 30.5 mL anhydrous ethanol, 2.4 mL deionized water, and 0.1 mL hydrochloric acid (HCl) with a concentration of 1 mol/L was prepared in a flask. This solution was stirred at room temperature for a duration of 30 minutes, followed by heating in a water bath maintained at \SI{60}{\degreeCelsius} for an additional 90 minutes. This sequential combination and controlled thermal procedure ensured the desired reaction conditions for the given synthesis process.

\begin{figure}[H]
  \centering
  \includegraphics[width=\linewidth]{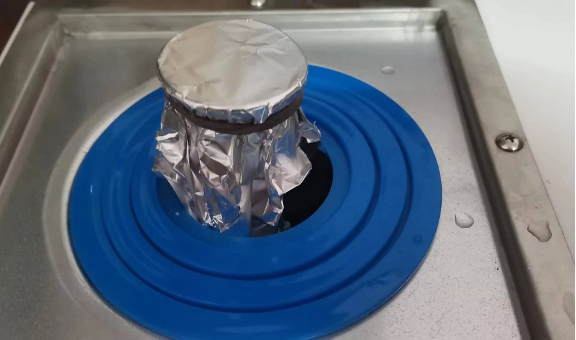}
\end{figure}

Take 10 mL of the above solution, add 44 mL of anhydrous ethanol and 1 mL of ammonia water with a concentration of 0.05 mol/L, stir at room temperature for 30 minutes, and place the whole liquid in a \SI{50}{\degreeCelsius} thermostat for aging and gelation for 96 hours.

\begin{figure}[H]
  \centering
  \includegraphics[width=\linewidth]{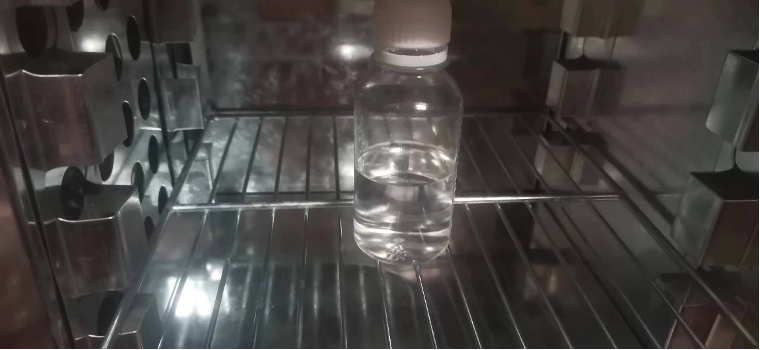}
\end{figure}
\begin{figure}[H]
  \centering
  \includegraphics[width=\linewidth]{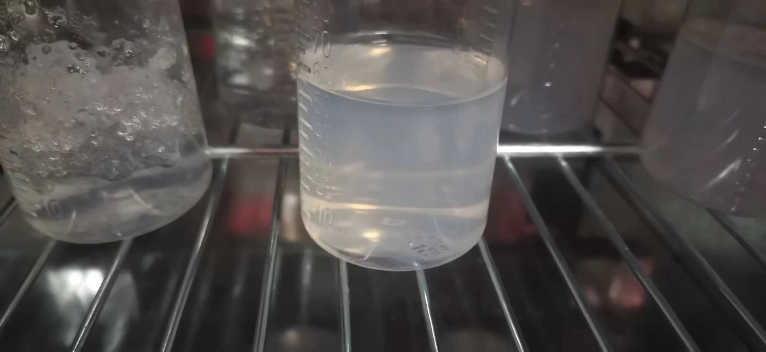}
\end{figure}
\begin{figure}[H]
  \centering
  \includegraphics[width=\linewidth]{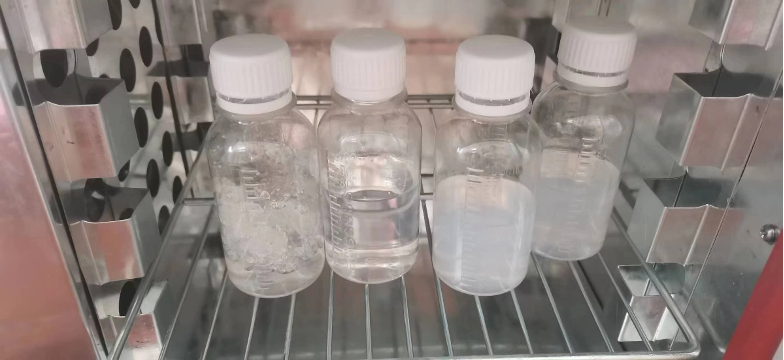}
\end{figure}

Soak the obtained gel in anhydrous ethanol for 3 hours first, then soak in n-hexane for 4 hours, then soak the gel in n-hexane containing 5\% v/v trimethylchlorosilane for 20 hours, and finally soak the gel in n-hexane for more than 2 hours.

All processes are carried out in a \SI{50}{\degreeCelsius} thermostat, and finally the gel is released by ultrasonic oscillation with 40 mL n-hexane and 30 mL acetylacetone (stabilizer), with ultrasonic time of 60 minutes, to obtain a flowable sol suitable for preparation.

\begin{figure}[H]
  \centering
  \includegraphics[width=\linewidth]{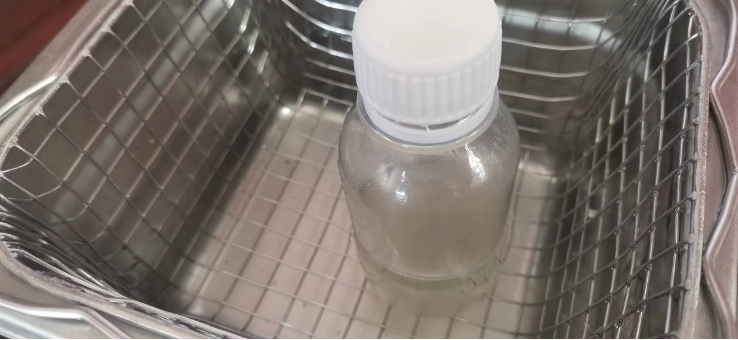}
\end{figure}

\subsection*{Preparation of aerogel thin films on solar cell surfaces}

\begin{figure}[H]
  \centering
  \includegraphics[width=\linewidth]{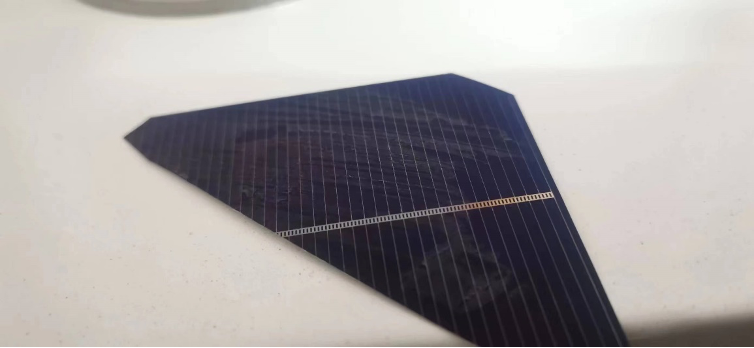}
\end{figure}
\begin{figure}[H]
  \centering
  \includegraphics[width=\linewidth]{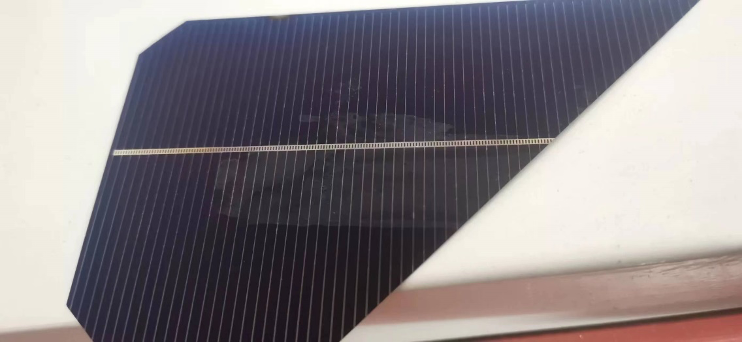}
\end{figure}

\begin{table*}[!b]
  \centering
  \tabcolsep = 20pt
  \begin{tabular}{ccccccc}
    \toprule
             & $V_{oc}$ & $V_{o}$@5$\Omega$ & $V_{o}$@4$\Omega$ & $V_{o}$@3$\Omega$ & $V_{o}$@2$\Omega$ & $V_{o}$@1$\Omega$ \\
    \midrule
             & 457      & 289                 & 250                 & 205                 & 150                 & 89                  \\
    w/o film & 450      & 286                 & 254                 & 213                 & 155                 & 90                  \\
             & 452      & 280                 & 238                 & 203                 & 153                 & 96                  \\
    \midrule
             & 468      & 312                 & 280                 & 219                 & 153                 & 95                  \\
    0.2mm    & 463      & 312                 & 278                 & 216                 & 149                 & 97                  \\
             & 466      & 310                 & 284                 & 220                 & 158                 & 95                  \\
    \midrule
             & 450      & 293                 & 265                 & 210                 & 147                 & 92                  \\
    0.4mm    & 460      & 294                 & 264                 & 215                 & 149                 & 96                  \\
             & 458      & 290                 & 263                 & 215                 & 139                 & 84                  \\
    \midrule
             & 428      & 265                 & 243                 & 194                 & 106                 & 76                  \\
    0.8mm    & 436      & 269                 & 241                 & 182                 & 112                 & 70                  \\
             & 430      & 263                 & 235                 & 186                 & 115                 & 68                  \\
    \bottomrule
  \end{tabular}
\end{table*}

\subsection*{Solar cell data measurement set up}

\subsubsection*{Preliminary testing set up}

  \begin{enumerate}

    \item  Connect the slide wire rheostat to the solar cell panel, first turn on the light source and measure the internal resistance of the solar cell;

    \item  Keep the light source unchanged, adjust the resistance value of the slide wire rheostat to be equal to the internal resistance of the solar cell panel;

    \item  Measure its output voltage and current.
  \end{enumerate}

  \begin{figure}[!ht]
    \centering
    \includegraphics[width=\linewidth]{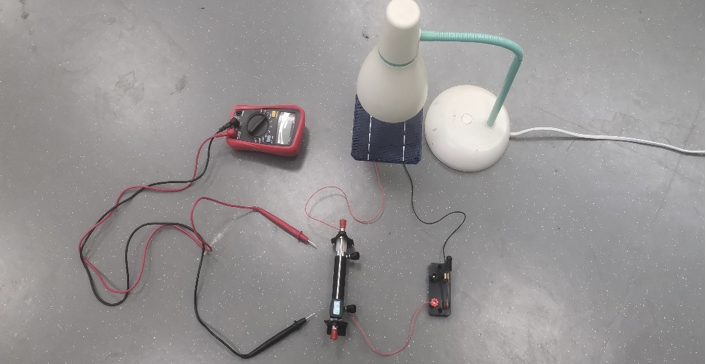}
  \end{figure}

  \subsubsection*{Selection of resistance}

  The selection of the resistance value for the variable resistor is critical for measuring the solar cell's maximal power output. In this particular configuration, the solar cell serves as a voltage supply once activated by light, undergoing a photoelectric conversion. Consequently, the current traversing the solar cell is equal to the current through the variable resistor, given their series connection. The relationships can be described by the following equations:

  \[I_s=\frac{V-V_o}{R_s}=\frac{V_o}{R_r}=I_R\]
  \[V_o=\frac{VR_r}{R_r+R_s}\]

  Where $I_{s}$ is the current through the solar cell, $I_{R}$ is the current through the resistor, $V$ is the total voltage, $V_{o}$ is the output voltage from the solar cell, $R_{s}$ is solar cell's internal resistance, and $R_{r}$ is the resistance of the variable resistor.

  The output power ($P_{o}$) is consequently given by:

  \[P_o=\frac{{V_o}^2}{R_r}=\frac{{(\frac{VR_r}{R_r+R_s})}^2}{R_r}=\frac{V^2R_r}{{(R_r-R_s)}^2+4R_rR_s}\]

  To optimize $P_{o}$, $R_{s}$ must be equal to $R_{r}$. Hence, the variable resistance must correspond to the internal resistance of the solar cell. The internal resistance was measured by directly connecting the multimeter to both ends of the solar cell, obtaining the corresponding resistance value.

  \subsection*{Preliminary measurement raw results}

  The table records the output voltage (V${}_{o}$) of solar cell connected to various resistance. The output power is calculated by:
  \[P_o=\frac{{V_o}^2}{R}\]

  Where $R$ is the resistance of the resistor.

\subsubsection*{IV curve measurement setup}

The optimal approach to electrical transportation measurement would necessitate the application of Kelvin four-terminal sensing. However, the irregular surface of the silica aerogel thin film precluded the possibility of direct wire bonding from the puck electrode to the sample surface. Consequently, a Printed Circuit Board (PCB) panel was employed as an intermediate, necessitating the selection of two-terminal sensing. The 3 x 5 mm sample piece was affixed to a puck using double-sided tape. Subsequently, two segments of coaxial cable were connected to the diagonal corners of the sample utilizing silver conductive paint, preceded by the removal of the oxide surface via sandpaper or a knife. The opposing ends of the coaxial cable were soldered to two conductive bands on the PCB panel. Thereafter, wire bonding (Shenzhen Wetel Semiconductor Equipment Co., Ltd., WL-2042) was employed to weld thin aluminum wire from the 0.5 mm diameter gold electrode to the corresponding conductive bands of the PCB panel. A supplementary channel was also established on the reverse sides of the puck electrodes, thereby enhancing the robustness of the setup. Finally, the puck was interfaced with a homemade user's bridge, which was in turn connected to two Keithley meters, completing the intricate yet precise assembly.

\begin{figure}[H]
  \centering
  \includegraphics[width=\linewidth]{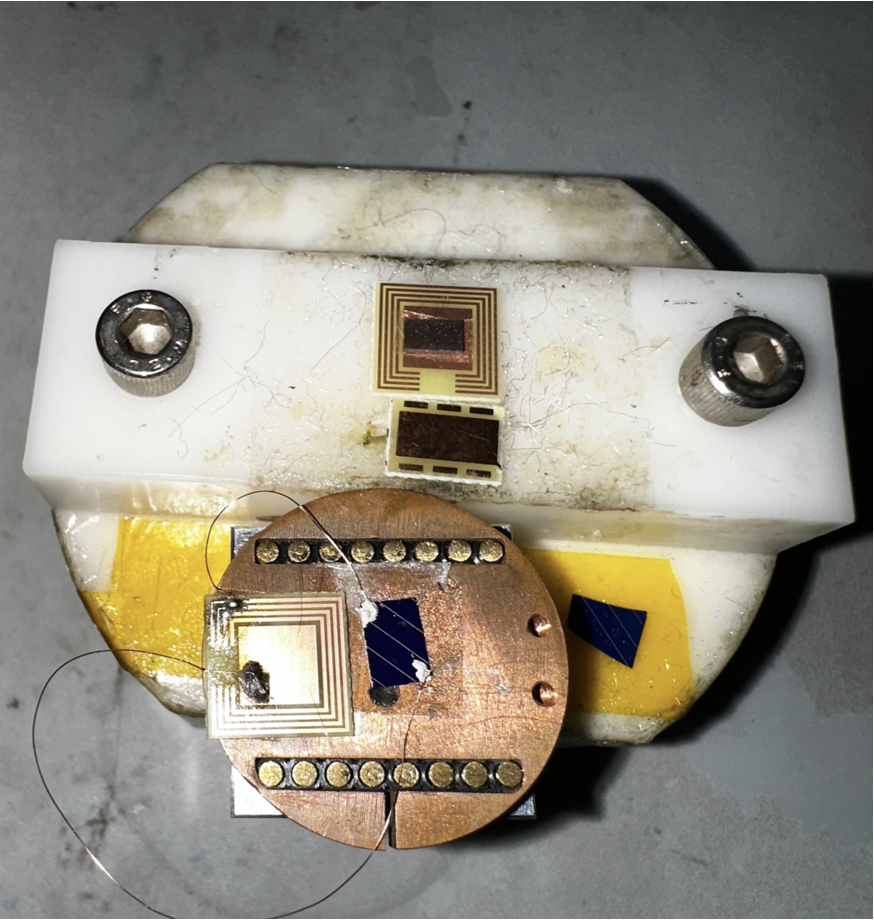}
\end{figure}

\end{document}